\documentclass[a4paper,dvips,12pt]{article}
\usepackage{epsf}
\usepackage{amsmath}
\numberwithin{equation}{section}

\newcommand{\be}{\begin{equation}}
\newcommand{\ee}{\end{equation}}
\newcommand{\ba}{\begin{eqnarray}}
\newcommand{\ea}{\end{eqnarray}}
\def\simlt{\mathrel{\lower2.5pt\vbox{\lineskip=0pt\baselineskip=0pt
             \hbox{$<$}\hbox{$\sim$}}}}
\def\simgt{\mathrel{\lower2.5pt\vbox{\lineskip=0pt\baselineskip=0pt
             \hbox{$>$}\hbox{$\sim$}}}}
\def\draftlabel#1{{\@bsphack\if@filesw {\let\thepage\relax
   \xdef\@gtempa{\write\@auxout{\string
     \newlabel{#1}{{\@currentlabel}{\thepage}}}}}\@gtempa
     \if@nobreak \ifvmode\nobreak\fi\fi\fi\@esphack}
      \gdef\@eqnlabel{#1}}
\def\@eqnlabel{}
\def\@vacuum{}
\def\draftmarginnote#1{\marginpar{\raggedright\scriptsize\tt#1}}
\def\draft{\oddsidemargin -.5truein
         \def\@oddfoot{\sl preliminary draft \hfil
         \rm\thepage\hfil\sl\today\quad\militarytime}
         \let\@evenfoot\@oddfoot \overfullrule 3pt
         \let\label=\draftlabel
         \let\marginnote=\draftmarginnote

\def\@eqnnum{(\theequation)\rlap{\kern\marginparsep\tt\@eqnlabel}%
\global\let\@eqnlabel\@vacuum}  }
\def\preprint{\twocolumn\sloppy\flushbottom\parindent 1em
         \leftmargini 2em\leftmarginv .5em\leftmarginvi .5em
         \oddsidemargin -.5in    \evensidemargin -.5in
         \columnsep 15mm \footheight 0pt
         \textwidth 250mmin      \topmargin  -.4in
         \headheight 12pt \topskip .4in
         \textheight 175mm
         \footskip 0pt

\def\@oddhead{\thepage\hfil\addtocounter{page}{1}\thepage}
         \let\@evenhead\@oddhead \def\@oddfoot{} \def\@evenfoot{}  }
\def\titlepage{\@restonecolfalse\if@twocolumn\@restonecoltrue\onecolumn
      \else \newpage \fi \thispagestyle{empty}\c@page\z@
         \def\thefootnote{\fnsymbol{footnote}} }
\def\endtitlepage{\if@restonecol\twocolumn \else  \fi
         \def\thefootnote{\arabic{footnote}}
         \setcounter{footnote}{0}}  
\catcode`@=12
\relax

\title{
\vspace*{-0.8cm}
\begin{flushright}
\normalsize{CERN--PH--TH/2004-217\\
\texttt{hep-th/0411032}}\\
\end{flushright}
\vspace{1cm}
\Large\textbf{Splitting Supersymmetry in String Theory}
\author{\large
{\bf I.~Antoniadis$^1$\footnote{On leave from CPHT 
(UMR CNRS 7644) Ecole Polytechnique, F-91128 Palaiseau},
S.~Dimopoulos$^{2}$}\\ \\
\emph{$^1$Department of Physics, CERN - Theory Division}\\
\emph{CH--1211 Geneva 23, Switzerland}\\
\emph{$^2$Physics Department, Stanford University}\\
\emph{Stanford, California 94309, USA}}}
\date{}
\begin{document}
\maketitle
\thispagestyle{empty}

\begin{abstract}
We point out that type I string theory in the presence of internal magnetic fields 
provides a concrete realization of split supersymmetry. To lowest order,
gauginos are massless while squarks and sleptons are superheavy. 
We build such realistic $U(3) \times U(2) \times U(1)$ models on stacks of 
magnetized D9-branes. Though not unified into a simple group, these theories 
preserve the successful supersymmetric relation of gauge couplings, 
as they start out with equal $SU(3)$ and $SU(2)$ couplings and the correct 
initial $\sin^2{\theta_W}$ at the compactification scale of 
$M_{\rm GUT}\simeq 2 \times 10^{16}$ GeV, and they have the 
minimal low-energy particle content of split supersymmetry. 
We also propose a mechanism in which the gauginos and higgsinos
are further protected by a discrete R-symmetry against gravitational corrections, 
as the gravitino gets an invariant Dirac mass by pairing with a member of a 
Kaluza-Klein tower of spin-3/2 particles.
In addition to the models proposed here, split supersymmetry offers novel 
strategies for realistic model-building. So, TeV-scale string models previously 
dismissed because of rapid proton decay, or incorrect $\sin^2{\theta_W}$, or 
because there were no unused dimensions into which to dilute the strength 
of gravity, can now be reconsidered as candidates for realistic split theories 
with string scale near $M_{\rm GUT}$, as long as the gauginos and 
higgsinos remain light.

\end{abstract}
\date

\newpage

\section{Introduction} \label{introduction}

Some recent developments challenge us to re-examine our preconceived notions of naturalness and our expectations for physics beyond the Standard Model at the LHC. First is the absence of any deviation from the Standard Model suggesting that, if there is new physics at a TeV, it appears to be fine-tuned at the per-cent level and does not comply with our notion of naturalness. Second, and most important, the cosmological constant problem (CCP) presents us with a fine-tuning much more severe than that of the gauge hierarchy problem (GHP). This raises the possibility that the mechanism which solves the CCP may also solve the GHP, and casts some doubts on all the mechanisms proposed so far to address the GHP (technicolor, low-energy supersymmetry (SUSY), low-scale strings, warping, little higgs), since none of them addresses the CCP problem. 

One concrete idea addressing the CCP is Weinberg's anthropic approach~\cite{weinberg} which postulated the existence of an enormous ``landscape'' of vacua, only a small fraction  of which have a vacuum energy small enough to  allow the formation of galaxies, which provide for a natural (and possibly necessary) habitat for observers such as ourselves. This approach has recently gained momentum because of the realization that string theory may have such a vast landscape of vacua ~\cite{bousso}. 
Such an environment may  drastically change what is a natural or likely theory. 
To see how this may happen, first recall that the standard measure of fine-tuning in SUSY theories is given by $ f_{standard} \sim {m_H^2 \over m_S^2}$, where $M_H \sim 100$ GeV is the Higgs mass and $m_S$ is the SUSY-breaking scale. 
Consider now a neighborhood in the landscape where the density of vacua increases with the scale of SUSY breaking proportional to $m_S^{2N}$ \cite{susskind}. Then, assigning equal a priori probability to each vacuum, the proper new measure of fine-tuning, which takes into account the ``entropy'' associated with the density of vacua, is $f_{new} \sim {m_H^2 \over m_S^2} \times  m_S^{2N}$. For $N > 1$, it thus favors large SUSY-breaking scale $m_S$. 

In such neighborhoods of the landscape low-scale SUSY is disfavored and, if we live in such a neighborhood, the simplest possibility is that we will discover the Standard Model (SM), rather than the supersymmetric Standard Model, at the LHC. This would
then account for why we have not seen any evidence for low-energy
supersymmetry,  at the expense of
giving up the two successes of the supersymmetric SM \cite{dg}: gauge
coupling unification~\cite{drw} and natural dark-matter (DM)
candidate~\cite{dg,gold}. A more interesting  possibility that
preserves these successes  is that  approximate chiral symmetries
protect the fermions of the supersymmetric SM  down to the TeV
scale ~\cite{savnim,noi,tutti,Arkani-Hamed:2004yi}. 
Actually, gauge coupling unification based on extrapolation 
of low-energy data to high energies is strictly speaking only 
an indirect indication of light gauginos and higgsinos, 
rather than of the full super-particle spectrum.

In these theories, the sparticle spectrum  is ``split" in two: (1) the scalars (squarks and
sleptons) that get a mass at the high-scale of supersymmetry
breaking $m_S$, which can be as large as the grand unification (GUT) scale, and (2)
 the fermions (gauginos and higgsinos) which are near the
electroweak scale and account for both gauge-coupling
unification and DM. The only light scalar in this theory is a
finely-tuned Higgs. Rather than the boring prediction that the
LHC will discover just the Higgs, these theories -- called Split
Supersymmetry -- predict gauginos and higgsinos at a TeV, maintain
the successes of the supersymmetric SM, and account for the
absence of any evidence for physics beyond the Standard Model, so far.

The main objective of this paper is to build models of split supersymmetry 
based on string theory.  It is clear that split supersymmetry offers novel, 
previously unavailable, strategies for realistic model-building, and some  
previously discarded classes of models can now be reconsidered.  
For example, classes of models of intersecting branes that were 
studied in the context of low-scale 
strings~\cite{Arkani-Hamed:1998rs,Antoniadis:1998ig}, and dismissed  
because of rapid proton decay or the value of the weak mixing angle 
will now be reconsidered, in section 6, and shown to contain good 
candidates for realistic theories. Some TeV-scale string models 
were also abandoned because of the absence of unused 
dimensions into which to dilute the strength of gravity~\cite{Uranga:2003pz}
should be reconsidered as candidates for realistic split 
theories with string scale near $2 \times10^{16}$ GeV, 
as long as the gauginos and higgsinos can remain light 
as a result of an approximate chiral symmetry. 

Another objective is  to build theories where the successful unification relation is preserved.  In split SUSY theories this is not a luxury but an essential ingredient, since unification is a fundamental phenomenological motivation for the ``split'' spectrum of these theories. This is a strong theoretical constraint, since it limits us to very economical fundamental theories with few relevant parameters in the gauge sector, small threshold corrections, minimal particle content, equal $SU(2)$ and $SU(3)$ couplings as well as the correct normalization of the weak mixing angle at the GUT scale ($\sin^2\theta_W=3/8$). 

Our paper is organized as follows.
In Section~\ref{framework} we discuss the theoretical framework, 
which is type I string theory with internally magnetized D9 branes. 
We show that the (tree-level) spectrum
of the resulting models is the one required by split supersymmetry.
In Section~\ref{unification}, we discuss the conditions that guarantee unification of
non-abelian gauge couplings and show that they can be naturally satisfied.
In Section~\ref{susybr}, we discuss the various mass scales and the
supersymmetry breaking in the gravity sector.
In Section~\ref{gauginos}, we propose mechanisms to keep gauginos
(and higgsinos) light in the presence of gravity.
In Section~\ref{models}, we study issues of model building
and present an explicit example of the SM embedding in the above framework,
with realistic particle spectrum, realizing the unification conditions and
predicting the correct weak mixing angle $\sin^2\theta_W=3/8$ at the GUT scale.
Finally, in Section~\ref{pheno}, we discuss some phenomenological consequences
and in particular constraints from gluino cosmology.

\section{The framework} \label{framework}

We start with type I string theory, or equivalently type IIB with 
orientifold planes and D-branes. Upon compactification 
in four dimensions on a Calabi-Yau manifold, 
one gets ${\cal N}=2$ supersymmetry in the bulk and 
${\cal N}=1$ on the branes. Moreover, various fluxes can be turned on, to 
stabilize part or all of the closed string moduli. We then turn on 
internal magnetic fields~\cite{Bachas:1995ik, Angelantonj:2000hi}, 
which, in the T-dual picture, amounts to 
intersecting branes~\cite{Berkooz:1996km, bi}. 
For generic angles, or equivalently for 
arbitrary magnetic fields, supersymmetry is spontaneously broken and 
described by effective D-terms in the four-dimensional (4d) 
theory~\cite{Bachas:1995ik}. In the weak field limit, 
$|H|\alpha'<1$, the resulting mass shifts are given by:
\be
\delta M^2=(2k+1)|qH|+2qH\Sigma\qquad ;\qquad k=0,1,2,\dots\, ,
\label{deltam}
\ee
where $H$ is the magnetic field of an abelian 
gauge symmetry, corresponding to a  Cartan generator of the higher 
dimensional gauge group, on a non-contractible 2-cycle of the 
internal manifold. $\Sigma$ is the corresponding projection of the 
spin operator, $k$ is the Landau level and  $q=q_L+q_R$ is the charge 
of the state, given by the sum of the left and right charges of the 
endpoints of the associated open string. We recall that the exact 
string mass formula has the same form as (\ref{deltam}) with $qH$ 
replaced by:
\be
qH\longrightarrow\theta_L+\theta_R\qquad ;\qquad 
\theta_{L,R}=\arctan(q_{L,R}H\alpha')\, ,
\label{stringdeltam}
\ee
where $\alpha'$ is the string Regge slope.
Obviously, the field theory expression 
(\ref{deltam}) is reproduced in the weak field limit.

To illustrate 
the physics, consider an effective six-dimensional (6d) theory 
compactified on a magnetized ``2-cycle". From the mass formula 
(\ref{deltam}), it follows that all charged scalars become massive, 
since the internal spin $\Sigma$ either vanishes (for six-dimensional 
scalars), or has eigenvalues $\pm 1$ (for 6d vectors). 
Actually, one of the two spin-1 helicities becomes tachyonic, 
reflecting the Nielsen-Olesen instability. This tachyon can be 
avoided, either  when several magnetic fields are turned on in more than one 
internal 2-cycles~\cite{Bachas:1995ik}, 
or in more realistic models with ${\cal N}=1$ supersymmetry 
in four dimensions. In the former case, one provides a positive 
contribution to its mass-squared (see below), while in the latter, one uses an
orbifold-type projection which reduces the supersymmetry from its
maximal value of the toroidal compactification to ${\cal N}=1$. On the other 
hand, fermions have four-dimensional (4d) chiral zero modes, since they 
have internal helicities $\Sigma=\pm 1/2$ and only one of the two leads to
a massless mode for $k=0$.  Note that neutral 
states with respect to the magnetized $U(1)$ generator are 
not affected and form ${\cal N}=1$ supermultiplets. In particular, all gauge 
bosons of the unbroken gauge group are accompanied by massless 
gauginos.

In the general case of a magnetic field pointed in several 
directions of the six-dimensional internal manifold, $H$ and 
$H\Sigma$ are replaced by ${\rm Tr}JH$ and ${\rm Tr}H\Sigma$, where 
$J$ is the antisymmetric ``identity" matrix with elements $+1(-1)$ 
above (below) the diagonal and zero everywhere else. For instance, 
when the internal manifold is a product of three factorized tori 
$\prod_{I=1}^3 T^2_{(I)}$, one has $H=\sum_I H_I$ and 
$H\Sigma=\sum_IH_I\Sigma_I$, where $\Sigma_I$ is the projection of 
the internal helicity along the $I$-th plane. For a ten-dimensional (10d) spinor,
its eigenvalues are $\Sigma_I=\pm 1/2$, while for a 10d 
vector $\Sigma_I=\pm 1$ in one of the planes $I=I_0$
and zero in the other two $(I\ne I_0)$. Thus, charged higher dimensional
scalars become massive, fermions lead to chiral 4d zero modes if all 
$H_I\ne 0$, while the lightest scalars coming from 10d vectors have masses
\be
M_0^2=\left\{\begin{matrix}
|qH_1|+|qH_2|-|qH_3|\cr 
|qH_1|-|qH_2|+|qH_3|\cr
-|qH_1|+|qH_2|+|qH_3|\cr
\end{matrix}\right.
\label{scalars}
\ee
Note that all of them can be made positive definite if all $H_I\ne 0$.
Moreover, one can easily show that if a scalar mass vanishes, 
some supersymmetry remains unbroken~\cite{Angelantonj:2000hi}.

The Gauss law for the 
magnetic flux implies that the fields $H_i$ are quantized in terms of 
the area of the corresponding 2-cycles $A_i$:\footnote{The
index $i$ becomes identical to $I$ above, when the 6d internal manifold
is a product of three factorized tori. In the general case, $i$ denotes
all possible two-cycles, even non-factorizable.}
\be
H_i={m_i\over n_i 
A_i}\, ,
\label{Hquant}
\ee
where the integers $m_i,n_i$ correspond 
to the respective magnetic and electric charges; $m_i$ is the 
quantized flux and $n_i$ is the wrapping number of the higher 
dimensional brane around the corresponding internal 2-cycle. 
For a rectangular torus of radii $R_1$ and $R_2$ in the directions
$X_1$ and $X_2$, the area is $A=R_1R_2$.
Open string propagation in magnetic fields has a T-dual representation
in terms of D-branes at angles. For instance, starting with a D$p$ brane  
on a magnetized rectangular torus and applying a T-duality in the direction 
$X_2$, $R_2\to\alpha'/R_2$, leads to a D$(p-1)$ brane wrapped on a 
direction forming an angle $\theta$ relative to the $X_1$ 
axis, given by the dual of the magnetic field: 
\be
H\alpha'\to\tan\theta={mR_2\over nR_1}\, .
\label{dual}
\ee
Thus, the integers $m$ and $n$ in (\ref{Hquant}) become the wrapping 
numbers around the $X_2$ and $X_1$ directions, respectively.

We consider now 
several abelian magnetic fields $H_i^a$ of different Cartan generators 
$U(1)_a$, so that the gauge group is a product of unitary factors 
$\prod_a U(N_a)$ with $U(N_a)=SU(N_a)\times U(1)_a$. In an 
appropriate T-dual 
representation, it amounts to consider several stacks of 
D6-branes intersecting in the three internal tori at angles 
determined by the magnetic fields according to (\ref{dual}). An 
open string with one end on the $a$-th stack has charge $\pm 1$ under 
the $U(1)_a$, depending on its orientation, and is neutral with 
respect to all others. Using the results described above, the 
massless spectrum of the theory falls into three sectors~\cite{bi}:
\begin{enumerate}
\item Neutral open strings ending on 
the same stack, giving rise to ${\cal N}=1$ gauge supermultiplets of gauge 
bosons and gauginos.
\item Doubled charged open strings from a 
single stack, with charges $\pm 2$ under the corresponding $U(1)$, 
giving rise to massless fermions transforming in the antisymmetric or 
symmetric representation of the associated $SU(N)$ factor. Their 
bosonic superpartners become massive. For factorized toroidal 
compactifications $(T^2)^3$, the multiplicities of chiral fermions 
are given by:
\ba
{\rm Antisymmetric}&:&\quad {1\over 2}\left(\prod_I 
2m_I^a\right)\left(\prod_J n_J^a+1\right)\nonumber\\
{\rm 
Symmetric}&:&\quad {1\over 2}\left(\prod_I 2m_I^a\right)\left(\prod_J 
n_J^a-1\right)
\label{dcmult}
\ea
where $a$ denotes the D-brane 
stack, $I$ is the label of the two-torus $T^2_{(I)}$, and $m_I^a, 
n_I^a$ are the integers entering in the expression of the magnetic 
field (\ref{Hquant}). For orbifolds or more general Calabi-Yau spaces, 
the above multiplicities may be further reduced by the corresponding 
supersymmetry projection down to ${\cal N}=1$.

In the degenerate case where 
a magnetic field vanishes, say, along one of the tori ($m_I^a=0$ for 
some $I$), there are no chiral fermions in $d=4$ dimensions, but the 
same formula with the products extending over the other two 
magnetized tori gives the multiplicities of chiral fermions in $d=6$. 
In this case, chirality in four dimensions may arise only when the 
last $T^2$ compactification is combined with some additional 
orbifold-type projection.
\item Open strings stretched between 
two different brane stacks, with charges $\pm 1$ under each of the 
corresponding $U(1)$s. They give rise to chiral fermions transforming 
in the bifundamental representation of the two associated unitary 
group factors. Their multiplicities, for toroidal compactifications, 
are given by:
\ba
(N_a,N_b)&:&\prod_I (m_I^a n_I^b+n_I^a 
m_I^b)
\nonumber\\
(N_a,{\overline N}_b)&:&\prod_I (m_I^a n_I^b-n_I^a 
m_I^b)\, .
\label{scmult}
\ea
As in the previous case, when a factor in the products of the above 
multiplicities vanishes, there are no 4d chiral 
fermions, but the same formula with the product extending over the 
other two magnetized tori gives the corresponding multiplicity of 
chiral fermions in $d=6$.
\end{enumerate}

As mentioned already above, 
all charged bosons are massive. Massless scalars can appear only when 
some supersymmetry remains unbroken. In case 2 of doubled charged 
strings from the same stack, the requirement of massless scalars is 
equivalent to unbroken supersymmetry on the corresponding brane stack. 
For toroidal compactifications, using the mass formula (\ref{scalars}), 
the condition for the $a$-th stack is
\be
\delta H_a\equiv
\epsilon_1H_1^a+\epsilon_2H_2^a+\epsilon_3H_3^a=0\, ,
\label{susya}
\ee
where $H_I^a$ is the magnetic field of $U(1)_a$ on the 
two-torus $T^2_{(I)}$, and $\epsilon_I$ are signs $\pm$ with one
at least different from the others
($(\epsilon_1,\epsilon_2,\epsilon_3)=(+,+,-)$ or $(+,-,+)$ or $(-,+,+)$).
In case 3 of strings ending on two different sets of branes, massless
scalars arise when one has unbroken supersymmetry locally, at the 
intersection. The generalization of the above condition is:
\ba
\delta H_{ab}\equiv\delta H_a-\delta H_b=0\, .
\label{susyab}
\ea
In the T-dual representation, condition (\ref{susyab}) involves
the relative intersection angles $(\theta_I^a-\theta_I^b)$, defined
as in eq.~(\ref{dual}).

It is now clear that the above framework leads 
to models with a tree-level spectrum realizing the idea of split 
supersymmetry. Embedding the Standard Model (SM) in an appropriate 
configuration of D-brane stacks, one obtains tree-level
massless gauginos while 
all scalar superpartners of quarks and leptons typically get masses at the  scale 
of the magnetic fields, whose magnitude is set by the compactification scale 
of the corresponding internal space. 

On the other hand, the condition to obtain a (tree-level)
massless Higgs in the spectrum implies that 
supersymmetry remains unbroken in the Higgs sector, leading to a pair 
of massless higgsinos, as required by anomaly cancellation. Note that 
since the Higgs doublet has the same quantum numbers with leptons, it 
is likely that lepton doublets have the same open string origin as 
the Higgs scalar, and thus, left-handed sleptons are also massless at 
the tree-level.

\section{Gauge coupling unification} \label{unification}

On general grounds, there are two conditions to obtain unification of
Standard Model gauge interactions, consistently with extrapolation of
gauge couplings from low-energy data using the minimal supersymmetric
SM spectrum. (i) Equality of the $SU(3)$ color and weak $SU(2)$ 
non-abelian gauge couplings and (ii) the correct prediction for the weak 
mixing angle $\sin^2\theta_W=3/8$ at the grand unification (GUT) scale.
On the other hand, a generic D-brane model using several stacks, as
described in the framework of the previous section, does not satisfy
either of the two conditions. Indeed, this framework was developed
in connection to the idea of low-scale 
strings~\cite{Arkani-Hamed:1998rs,Antoniadis:1998ig}, where the concept of
unification is radically different from conventional GUTs.
In this section, we study precisely the general requirements for 
satisfying the first of the above two conditions, namely
natural unification of non-abelian gauge couplings.
The second condition is more involved and model dependent, 
since it is related with the particular hypercharge embedding and
will be discussed in section~\ref{models}.

The four-dimensional non-abelian gauge coupling $\alpha_{N_a}$ of the 
$a$-th brane stack is given by:
\ba
{1\over \alpha_{N_a}}={V^a\over 
g_s}\prod_I |n_I^a|\sqrt{1+(H_I^a\alpha')^2}\, ,
\label{ga}
\ea
where $g_s$ is the string coupling and $V^a$ the compactification volume 
in string units of the internal space of the $a$-th brane stack. 
The presence of the wrapping numbers $|n_I^a|$ can be understood from 
the fact that $|n_I^a|V^a_I$ is the effective area of the 2-torus $T^2_{(I)}$
wrapped $n_I^a$ times by the brane, and $V^a=\prod_IV^a_I$.
The additional factor in the square root follows from the non-linear
Dirac-Born-Infeld (DBI) action of the abelian gauge field, 
$\sqrt{\det (\delta_{ij}+F_{ij}\alpha')}$, which in the case of two dimensions
with $F_{ij}=\epsilon_{ij}H$, it is reduced to $\sqrt{1+(H\alpha')^2}$.
Obviously, the expression (\ref{ga}) holds at the compactification scale, 
since above it gauge couplings receive 
important corrections and become higher dimensional. 
Finally, the gauge couplings of the associated abelian factors, in our 
convention of $U(1)$ charges, are given by
\be 
\alpha_{_{U(1)_a}}={\alpha_{N_a}\over{2N_a}}\, . 
\label{gua}
\ee
Here, non-abelian generators are normalized according to 
${\rm Tr}T^aT^b=\delta^{ab}/2$. 

From equation (\ref{ga}), it follows that 
unification of non-abelian gauge couplings holds if (i) $V^a$ and 
(ii) $\prod_I|n_I^a|$ are independent of $a$, while (iii) the 
magnetic fields are either $a$-independent as well, or they are much 
smaller than the string scale. 
\begin{itemize}
\item The first 
condition (i) is automatically satisfied for D9-branes, since then
$V^a=V$, the total volume of the six dimensional internal manifold. 
\item The second condition (ii) is satisfied for a large class 
of models with $|n_I^a|=1$, which is the point particle field theoretic value of 
Dirac quantization for magnetic fields (no multiple brane wrapping). 
Actually, this value follows also from eq.~(\ref{dcmult}), by 
requiring the absence of chiral fermions transforming in the 
symmetric representations of the non-abelian groups, {\em i.e.} no 
chiral $SU(3)$ color sextets and no weak $SU(2)$ 
triplets.\footnote{The vanishing of the 
multiplicity (\ref{dcmult}) is also realized when some $m_I^a=0$, 
which is a trivial solution since in this case the corresponding 
magnetic field vanishes.}
\item The third condition (iii) of weak 
magnetic fields is more quantitative. Allowing for $1\%$ error in the 
unification condition at high scale, one should have 
$H_I^a\alpha'\simlt 0.1$. From the quantization condition 
(\ref{Hquant}), this implies that the volume $V\simgt 10^3$ for three 
magnetized tori, which is rather high to keep the theory weakly 
coupled above the compactification scale. Indeed, eq.~(\ref{ga})
gives a string coupling $g_s$ of order ${\cal O}(10)$ for gauge couplings
$\alpha_{N_a}\simeq 1/25$ at the unification scale. On the other hand, 
for one or two magnetized tori one obtains $V\simgt 10-10^2$, 
which is compatible with a string weak 
coupling regime $(g_s\sim 0.1-1)$. Of course this discussion
should be taken with caution, because there is an uncertainty
in the relation of $g_s$ with the string loop expansion parameter.
Here, we were conservative and defined it as in a 4d gauge theory.
In a 10d theory however, there may be additional powers of $2\pi$
which would improve significantly perturbativity~\cite{ah}.

Actually, the condition of weak magnetic fields can 
be partly relaxed in some direction, by requiring the absence of 
chiral antiquark doublets in the spectrum. Indeed eq.~(\ref{scmult}),
for open strings stretched between the strong $SU(3)$ and 
weak $SU(2)$ interactions brane stacks, 
implies the vanishing of one of the factors in the product. 
This leads to the equality of the ratio $m_I^a/n_I^a$ for the 
two stacks and for some $I$, and thus, to the equality of the two 
corresponding magnetic fields via eq.~(\ref{Hquant}).\footnote{This 
argument is true only when the $U(1)$ accompanying the weak 
interactions brane stack participates in the hypercharge combination. 
Otherwise, quark anti-doublets are equivalent to quark doublets 
(see example in section~\ref{models}).} As a result, 
the condition of perturbativity is weakened and becomes 
possible even in the case of three factorized magnetized tori.
\end{itemize}

Note that in the T-dual representation of 
intersecting D6-branes, the unification conditions discussed above
appear less natural. 
In the expression (\ref{ga}) of gauge couplings, the numerator
($V^a$ times the product) 
is replaced by the volume of the 3-cycle around which the 
D6 brane wraps. For instance, in the case of three factorized 
rectangular tori of radii $R^I_1$ and $R^I_2$ in string units, it is 
given by $\prod_I\sqrt{(n_I^aR_1^I)^2+(m_I^aR_2^I)^2}$. 
The same unification conditions then hold in this context, 
with the requirement of weak magnetic field replaced 
by the requirement of small angle, which is equivalent to
the inequality $R_2^I<<R_1^I$ (see eq.~(\ref{dual})).

The above analysis concerns mainly the QCD and 
$SU(2)_L$ gauge couplings $\alpha_3$ and $\alpha_2$. 
The case of hypercharge is more subtle since 
it can be in general a linear combination of several $U(1)$s coming 
from different brane stacks. 
In section~\ref{models}, for the purpose of illustration, 
we present an explicit example with the correct prediction of the
weak mixing angle. It is based on a minimal Standard 
Model embedding in three brane stacks with the hypercharge being a 
linear combination of two abelian factors. This provides an existence
proof that can be generalized in different constructions.
We notice for instance that in a class of supersymmetric
models with four brane stacks, the equality of the two non-abelian
couplings $\alpha_2=\alpha_3$ implies the value $3/8$ for
$\sin^2\theta_W$ at the unification scale~\cite{Blumenhagen:2003jy}.

\section{Mass scales and supersymmetry breaking}
 \label{susybr}

The supersymmetry breaking scale $m_{\rm S}$ 
on the brane stacks is given by
the lightest charged scalar masses (\ref{scalars}), or equivalently by 
$\delta H_a=\sum_I\epsilon_I H_I^a$ of eq.~(\ref{susya}), in the
weak field limit. In the case of strong magnetic fields, of order
of the string scale, $H_I^a$ should be replaced by the angles
$\theta_I^a$ according to eqs.~(\ref{stringdeltam}) and (\ref{dual}).
For magnetic fields in more than one internal planes, 
$m_{\rm S}$ can therefore be smaller 
than their magnitude, and consequently from the 
corresponding compactification scales (\ref{Hquant}).
Similarly, on brane intersections, the supersymmetry breaking 
scale is given by the differences $\delta H_{ab}$ of eq.~(\ref{susyab}),
and thus, can be again smaller than $H_I^a$ and the 
compactification scales.

Let us now discuss the various mass scales.
To preserve gauge coupling unification, the (non-gravitational part of the) theory must remain 4-dimensional up to the unification scale.  So  the compactification scale (actually the smallest, if there are several)
must be no smaller than the unification energy, 
$M_{\rm GUT}\simeq 10^{16}$ GeV, and we will take them to be of the same order. Above the compactification scale, gauge interactions
acquire a higher dimensional behavior.
So, to keep the theory weakly coupled, the string
scale $M_s\equiv{\alpha'}^{-1/2}$ should be close to the 
compactification scale and therefore to $M_{\rm GUT}$.
Moreover, as we discussed in the previous section, 
to ensure that corrections to the unification of gauge couplings
are within 1\%, the magnetic fields should be weak.
It follows that the string scale should be roughly a factor of 3 
higher than the compactification scale, 
\be
M_s\simgt 3\, M_{\rm GUT}\, .
\label{Ms}
\ee

On the other hand, as we pointed out above, $m_{\rm S}$
can be lower than $M_{\rm GUT}$. Although much lower
values require an apparent fine tuning of radii, such a 
tuning is technically natural since the supersymmetric 
point $m_{\rm S}=0$ is radiatively stable.
One can therefore treat $m_{\rm S}\sim|\delta H|^{1/2}$ 
as free parameter and drop for simplicity the brane stacks 
dependent index in $\delta H$.

All scalar masses are of order $|\delta H|^{1/2}$ except for
those coming from supersymmetric sectors, which are vanishing to 
lowest order, such as the Higgs and possibly the slepton doublets.
The latter are expected to acquire masses from one loop 
corrections, proportional to $|\delta H|^{1/2}$ but suppressed 
by a loop factor. 
Note that off diagonal elements of the $2\times 2$ Higgs 
mass matrix, usually denoted by $B\mu$, should also 
be generated at the same order as the diagonal elements, 
in the absence of a Peccei-Quinn (PQ) symmetry. For 
high $\delta H$, a fine tuning between $B\mu$ and the diagonal 
elements is then required to ensure a light higgs. 

It remains to discuss the corrections to gaugino and 
higgsino masses, $m_{1/2}$ and $\mu$, which are vanishing at the 
tree-level. In the absence of gravity, they are both protected by an 
R-symmetry. Actually, higgsino masses are protected in addition by a 
PQ symmetry which must be broken in order to generate a $B\mu$ 
mixing term in the Higgs mass matrix, as we argued above. 
Then, a $\mu$-term can be generated via $B\mu$, 
or directly using the PQ symmetry breaking, 
if R-symmetry is broken. Indeed,
R-symmetry is in general broken in the gravitational sector 
by the gravitino mass $m_{3/2}$ and thus, in the presence 
of gravity, $m_{1/2}$ and $\mu$ are not anymore protected.

Thus, the study of fermion masses requires some knowledge of
supersymmetry breaking in the gravity sector, which has been
ignored up to now. A related issue is the cancellation of the
cosmological constant between brane and bulk contributions,
in order to maintain the flat space background.
The brane contribution comes from the supersymmetry breaking
due to the magnetic field and scales as $(\delta H)^p$, in string units,
where the power $p$ depends on the number ${\cal N}$ of bulk 
supersymmetries broken by $\delta H$. For the maximal value
of ${\cal N}=4$ it was found that $p=3$~\cite{Bachas:1995ik}, 
while for ${\cal N}=1$ we expect that $p=1$, 
or more precisely $(\delta H)H$ from the
form of the DBI action (\ref{ga}). 

The vanishing of the vacuum energy implies that an additional 
source of supersymmetry breaking should probably be introduced
in the closed string sector (bulk).
The corresponding dominant bulk contribution to the cosmological
constant is proportional in general to $m_{3/2}^2\Lambda^2$, 
with $\Lambda$ the ultraviolet (UV) cutoff.
Combining the brane and bulk contributions, one obtains
\be
m_{3/2}\sim {|H|^{1/2}\over \Lambda}\times |\delta H|^{1/2}\, .
\label{mgrav}
\ee
Thus, for $\Lambda\simeq M_s$ the gravitino mass is
in general of the same order as the scalar masses,
while for $\Lambda\simeq M_P$, the 4d Planck mass, it is 
about three orders of magnitude lower.

In the following, we will consider for concreteness a source of
bulk supersymmetry breaking via Scherk-Schwarz (SS) ~\cite{SS} boundary
conditions along a ``gravitational" interval $S^1/Z_2$ 
of length $\pi R$~\cite{Antoniadis:1996hk}. This interval
can be identified either with the eleventh dimension of M-theory~\cite{Horava:1996ma}, or
with some internal orientifold direction of type I string theory, transverse
to all ``observable" brane stacks where the Standard Model is
localized~\cite{Antoniadis:1998ki}. The gravitino mass is then given by:
\be
m_{3/2}={\omega\over R}\, ,
\label{mgravitino}
\ee
where $\omega\in [0,1]$ is the parameter of the SS deformation.
It originates from the boundary conditions of the five dimensional 
fields which are periodic up to a phase 
of a symmetry transformation~\cite{SS}. 
The latter can be parametrized as $e^{2i\pi\omega/R}$ 
and corresponds to a discrete rotation in the internal compactified 
space, in order to give a mass to the gravitino. Therefore,
$\omega$ is quantized and equals $1/N$ for $Z_N$.

Using now the usual relations that express the 4d Planck mass 
$M_P$ and the gauge coupling at the unification scale $\alpha_G$ in
terms of the string parameters (scale, coupling and compactification
sizes), one finds upon eliminating the string 
coupling~\cite{Antoniadis:1996hk, Antoniadis:1998ig}:
\be
R^{-1}={2\over\alpha_G^2}{M_s^3\over M_P^2}V^{-1}\, ,
\label{RSS}
\ee
where $V$ is the internal compactification volume (in string units) 
of all SM branes. Substituting the values discussed above for 
$M_s$ and $V$, one finds $R^{-1}\simeq{\cal O}(10^{13}-10^{14})$ 
GeV. Following our previous discussion on the cancellation of
vacuum energy, when combining the SS bulk supersymmetry 
breaking with the brane magnetic fields, one expects that
$|\delta H|^{1/2}$ should also be of the same order $10^{13}$ GeV.
On the other hand, because of the uncertainty in the value of the
relevant UV cutoff in eq.~(\ref{mgrav}), the scalar masses could be
either significantly higher, of order of the unification scale,
or lower, of order of $10^{10}$ GeV for $\Lambda\simeq R^{-1}\simeq m_{3/2}$,
as was argued in the context of SS compactifications~\cite{Antoniadis:1990ew}.
Thus, in section~\ref{pheno}, we will study the phenomenology of the
whole range of scalar masses $m_S\simeq 10^{10}-10^{16}$ GeV.

\section{Light gaugino masses} \label{gauginos}

In the presence of R-symmetry, gluinos can only get a Dirac mass 
by pairing up with other color octet fermions, which spoils gauge 
coupling unification. So, gluinos must either be massless, 
which is phenomenologically strongly disfavored, or get an 
R-breaking Majorana mass. The latter requires a source for R-breaking, 
and would also permit, in combination with PQ breaking, 
the generation of higgsino masses.

One possible source of R-symmetry breaking is the Majorana 
mass for the gravitino. Such a mass is always present, 
as a result of canceling the tree-level cosmological constant, 
in theories where there is an energy regime in which 4d supergravity holds. 
A second possibility is that there is no such an energy regime, 
the gravitino gets an R-preserving Dirac mass by pairing up 
with another spin-3/2 fermion, and R-symmetry is broken 
spontaneously by a dynamical condensate. In this section, we will 
consider both possibilities, beginning with the first.

The first possibility has already been studied in some detail 
in the effective field theory~\cite{Antoniadis:1997ic,savnim,Arkani-Hamed:2004yi}. 
Once R-symmetry, as well as supersymmetry, is broken through 
the Majorana gravitino mass, the gauginos can get a mass 
in a number of ways. One is anomaly mediation~\cite{Randall:1998uk}, 
whose leading contribution
can be adequately suppressed  to allow for light gauginos in the presence of heavy gravitinos~\cite{Antoniadis:1997ic,Arkani-Hamed:2004yi}.

In string theory, gaugino masses mediated from closed string radiative corrections 
have been studied recently and shown to be generated at lowest order by 
string diagrams of ``one and a half" loop (``genus" 3/2)~\cite{Antoniadis:2004qn}. 
They contain for instance one handle and one boundary. 
It turns out that for generic compactifications and 
supersymmetry breaking mechanism, the resulting gaugino 
masses are proportional to the gravitino mass for small 
$m_{3/2}$ compared to the string scale: $m_{1/2}\sim\alpha^2 m_{3/2}$ 
with $\alpha$ the corresponding gauge coupling. 
Apart from the power of gauge coupling, this 
result is similar to the contribution of anomaly mediation~\cite{at}.

Suppression of the anomaly mediation contribution 
in radiative corrections may arise as follows.
A generic contribution to the gaugino mass
involves gravity and gauge loops and should contain a 
gravitino mass insertion that brings one power of $m_{3/2}$.
From an effective field theory analysis, one expects that the 
dominant contribution of each gravity loop is proportional to 
$\Lambda^2/M_P^2$, with $\Lambda$ the UV cutoff, 
since each gravitational vertex brings an inverse power of 
Planck mass $M_P$ and the loop is quadratically divergent. 
Moreover, gauge loops do not modify this power counting.
If the UV cutoff is set up by the Planck scale, $m_{1/2}$ would be
proportional to $m_{3/2}$.
However, in special models with supersymmetry breaking via 
Scherk-Schwarz compactifications, one expects a UV cutoff 
set by the compactification scale~\cite{Antoniadis:1990ew}, 
in which case the dominant contribution comes 
from one gravitational loop, leading to 
$m_{1/2}\propto m_{3/2}^3/M_P^2$~\cite{Antoniadis:1997ic,savnim}.
Thus, gauginos (and higgsinos) become light around the TeV 
scale, for instance when 
$m_{3/2}\sim|\delta H|^{1/2}\simeq 10^{13-14}$ GeV 
as we argued above.

On the other hand, in the string theory analysis,
it was found that for orbifold compactifications the 
corrections to $m_{1/2}$ are exponentially suppressed for small 
gravitino mass at the lowest non-trivial order 
of ``genus" 3/2~\cite{Antoniadis:2004qn}.
This is an indication that these models are indeed examples
of quantum gravitational suppressed anomaly mediation, 
which may be checked by going to the next non-trivial order.

We now present a different mechanism to protect gaugino 
masses in the presence of gravity, based on symmetry.
It involves theories in which there is no 4d supergravity energy regime, 
the gravitino gets an R-preserving Dirac mass by pairing up with 
another spin-3/2 fermion, and therefore does not feed a mass to the gluino. 
The breaking of R-symmetry, which is necessary to give masses to the 
gluinos and higgsinos, can subsequently occur via a dynamical condensate.

We begin from the observation that the lowest order
perturbative correction to $m_{1/2}$ is exactly vanishing for
the case of $Z_2$ SS deformation with $\omega=1/2$ (see 
eq.~(\ref{mgravitino}))~\cite{Gherghetta:2001sa, Antoniadis:2004qn}. 
Indeed, here we will argue that the 
usual SS compactification with a $Z_2$ shift in a 
direction transverse to the brane 
leaves unbroken a generalized R-type symmetry, which 
guarantees the vanishing of gaugino masses in the full theory. 
We will use the effective field theory description of the whole tower 
of Kaluza-Klein (KK) excitations for a generic SS compactification
on a circle of radius $R$~\cite{Bagger:2001ep}. 
A massless five-dimensional (5d)
spinor $\psi$, with SS boundary conditions twisted by the phase
$e^{2i\pi\omega/R}$ around the circle, gives rise in four 
dimensions to the following $2\times 2$ mass matrix for 
each pair of KK levels $|n>$ and $|-n>$~:
\ba
{1\over R}
\left(\psi^L_n\ 
\psi^R_n\right)
\left(\begin{matrix}
\omega&n\cr
n&\omega\cr
\end{matrix}\right)
\left(\begin{matrix}
\psi^L_n\cr
\psi^R_n\cr
\end{matrix}\right)\quad 
;\quad n=0,1,2\dots\, ,
\label{KKmass}
\ea
where $\psi^{L,R}_n$ is 
the left (L) and right (R) component of the $n$-th KK excitation of 
the fermion. The diagonal element $\omega$ arises from
the SS deformation. The eigenvalues of the above mass 
matrix are $(\omega\pm n)/R$, reproducing the familiar shift of the 
KK number. 

Here, we consider a SS direction which is 
transverse to the brane stack, so that gauginos are not affected 
and supersymmetry remains unbroken 
on the branes in the presence of the SS deformation. 
The only source of supersymmetry breaking on the branes comes
from the magnetic fields on their world volume, which give 
masses to all charged scalars, as we described previously. 
Thus, in this case, the SS direction is not a circle but an interval 
$S^1/Z_2$, with $Z_2$ being the inversion of the extra coordinate.
Its action on the KK spectrum consists of sending $|n>\to|-n>$, 
while at the same time acts on 4d fermion chiralities:
left-hand components are invariant and right-handed change sign.
As a result, the $Z_2$ projection on the KK spectrum of a 5d spinor
keeps the left-handed symmetric and right-handed antisymmetric
combinations of states $(|n>_{_L}+|-n>_{_L})/\sqrt{2}$ 
and $(|n>_{_R}-|-n>_{_R})/\sqrt{2}$, having cosine and 
sine wave functions, respectively. Keeping the same notation
$\psi^L_n$ and $\psi^R_n$ for these $Z_2$ invariant combinations,
the fermion mass terms can be easily deduced from the
expression (\ref{KKmass}) and take the form:
\ba
{\omega\over R}\psi^L_0\psi^L_0+{1\over R}\sum_{n\ge 1}
\left(\psi^L_n\ \psi^R_n\right)
\left(\begin{matrix}
\omega&n\cr
n&\omega\cr
\end{matrix}\right)
\left(\begin{matrix}
\psi^L_n\cr
\psi^R_n\cr
\end{matrix}\right)\, .
\label{KKmassorb}
\ea 

For a generic SS deformation, corresponding for instance to a 
$Z_N$ shift with $\omega=1/N$ and $N>2$, a simple inspection of 
eq.~(\ref{KKmassorb}) shows that there is a tower of Majorana 
masses for the gravitino KK modes,
that break the R-symmetry of global supersymmetry.
Thus, gauginos are expected to acquire masses through
gravitational radiative corrections. Moreover, despite the
quantization of $\omega$, one can define three energy regimes
in the gravity sector. A low-energy 4d non-supersymmetric region
below the lightest KK gravitino mass $\omega/R$, a 5d supergravity
region at energies higher than the compactification scale $E>>1/R$,
and an intermediate regime at energies $\omega/R<E<1/R$,
where one can define a 4d ${\cal N}=1$ spontaneously broken 
supergravity. The latter can be obtained by integrating out all
heavier KK excitations with $n>1$ and describes the physics of
the gravitino ``zero mode" $n=0$.

This general picture breaks down in the $Z_2$ case $\omega=1/2$,
due to a new pairing that arises in the KK spectrum.
The ``zero mode" becomes degenerate with the lightest eigenstate 
of the $2\times 2$ mixing matrix for the first KK excitation $n=1$, 
with mass eigenvalue $1/2R$. This degeneracy continues 
similarly to all KK levels; the heaviest eigenstate at level $n$
with mass eigenvalue $(n+1/2)/R$, becomes degenerate with the 
lightest eigenstate at level $n+1$. 
Thus, all masses can be rewritten in a Dirac type form and one
can define a new unbroken R-symmetry that keeps gauginos 
on the transverse branes massless. Note also that in this
case, there is no intermediate energy regime
where one can define a 4d ${\cal N}=1$ supergravity, since
after the SS deformation, the 4d gravitino zero mode is degenerate
with another state coming from its $n=1$ KK excitation.
Including this extra spin-3/2 state in the effective theory, one
should also include its degenerate $n=1$ companion at the
symmetric phase, which however, after the SS deformation, 
becomes degenerate with the lightest eigenstate from the
next level $n=2$, and so on. In the effective supergravity,
one should therefore include the whole KK tower and the
intermediate energy regime is lost.  

The low-energy 4d non-supersymmetric region 
without any gravitino mode has obviously
a chiral symmetry associated to the massless gauginos. 
On the other hand, in order to describe the generalized 
R-symmetry in the presence of gravity, one has to go directly 
to the high energy 5d regime with the whole KK gravitino tower.
This phenomenon provides the first example of massive 
supergravity coupled to an exact supersymmetric gauge sector
(but non-supersymmetric chiral matter) 
that survives in the quantum theory.
As we mentioned earlier, the R-symmetry can be broken 
spontaneously by appropriate dynamics
within the effective field theory at much lower 
energies and generate gaugino and higgsino masses close to the 
electroweak scale, preserving the unification of gauge 
couplings. In this case, the corresponding breaking scale
is an extra parameter that requires separate dynamics.

\section{Model building} \label{models}

In this section, we present an explicit Standard Model embedding, 
in a minimal set of three brane stacks which has a realistic particle content,
satisfies the conditions of unification of strong and weak interactions
and predicts the correct weak angle $\sin^2\theta_W=3/8$ at the
unification scale. This model illustrates our general framework
and provides an explicit example where several problems can be
addressed and many general phenomenological consequences can
be discussed.

Model building with intersecting branes has been extensively 
studied in the recent literature, mainly in the context of low-scale string 
models~\cite{Arkani-Hamed:1998rs, Antoniadis:1998ig} 
(see for example ref.~\cite{Uranga:2003pz} and references therein). 
According to the general analysis of ref.~\cite{Antoniadis:2000en}, the 
SM  embedding requires usually four stacks of branes, 
the color $U(3)$, the weak $U(2)$, together with two abelian ones. The 
hypercharge is in general a linear combination of the four $U(1)$s, 
while the remaining three orthogonal combinations are usually broken 
by anomalies to their global counterparts corresponding to the baryon 
and lepton numbers and a Peccei-Quinn symmetry. Moreover, the 
value of the weak angle when $\alpha_3=\alpha_2$ is in general 
different from $3/8$, which in any case is not a desired value when 
the string scale is at low energies. Here, we will focus on a 
particular model that was dropped from the analysis of 
refs.~\cite{Antoniadis:2000en} because, although minimal 
and very economic, it was not appropriate for low string scale. 
Its two main defects were the value of the weak angle and the absence 
of baryon number as a symmetry to guarantee proton stability.

\begin{figure}
\centerline{
\epsfxsize=4.4cm
\epsfbox{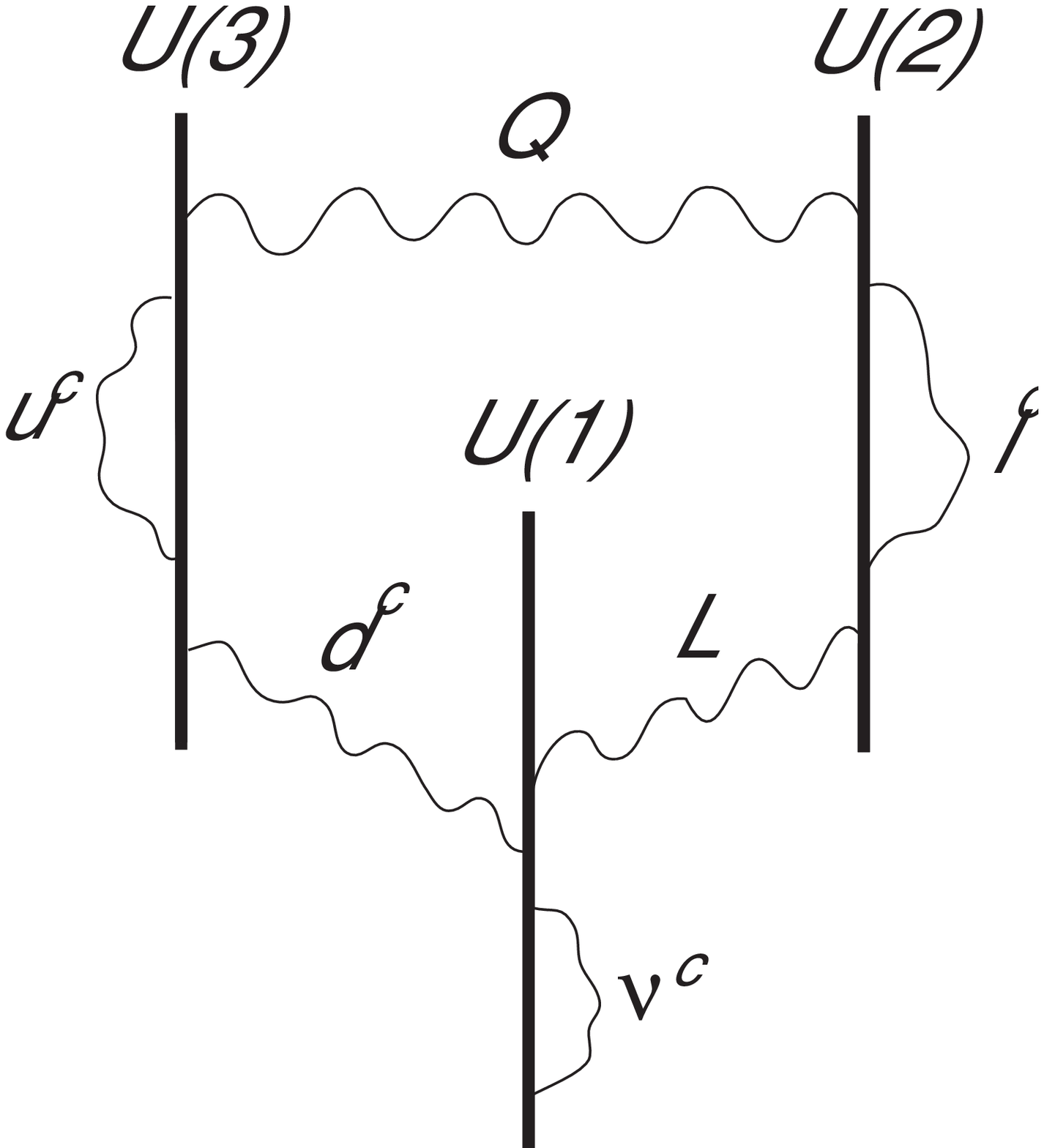}~~~\epsfxsize=4.4cm
\epsfbox{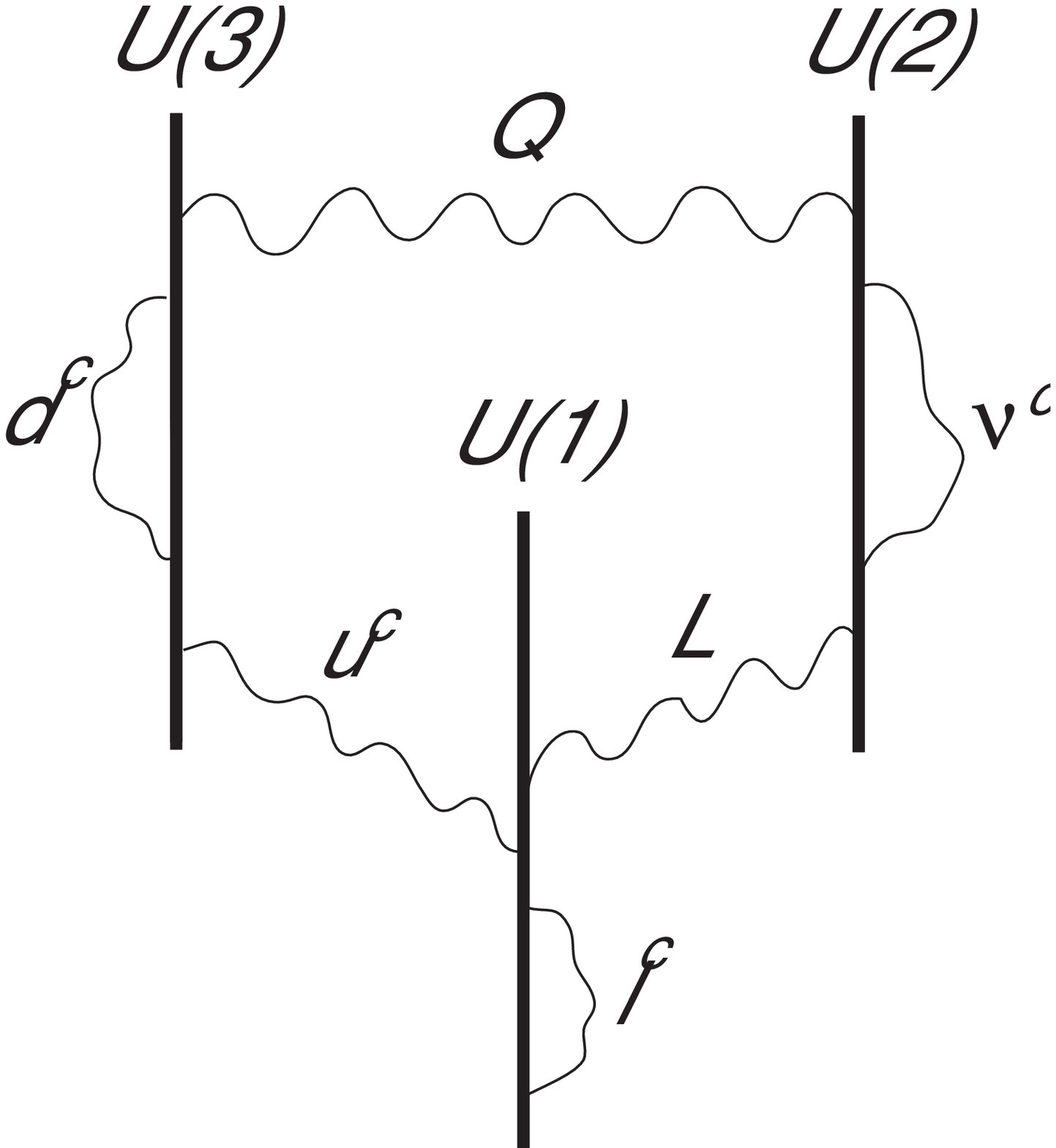}~~~\epsfxsize=3.9cm
\epsfbox{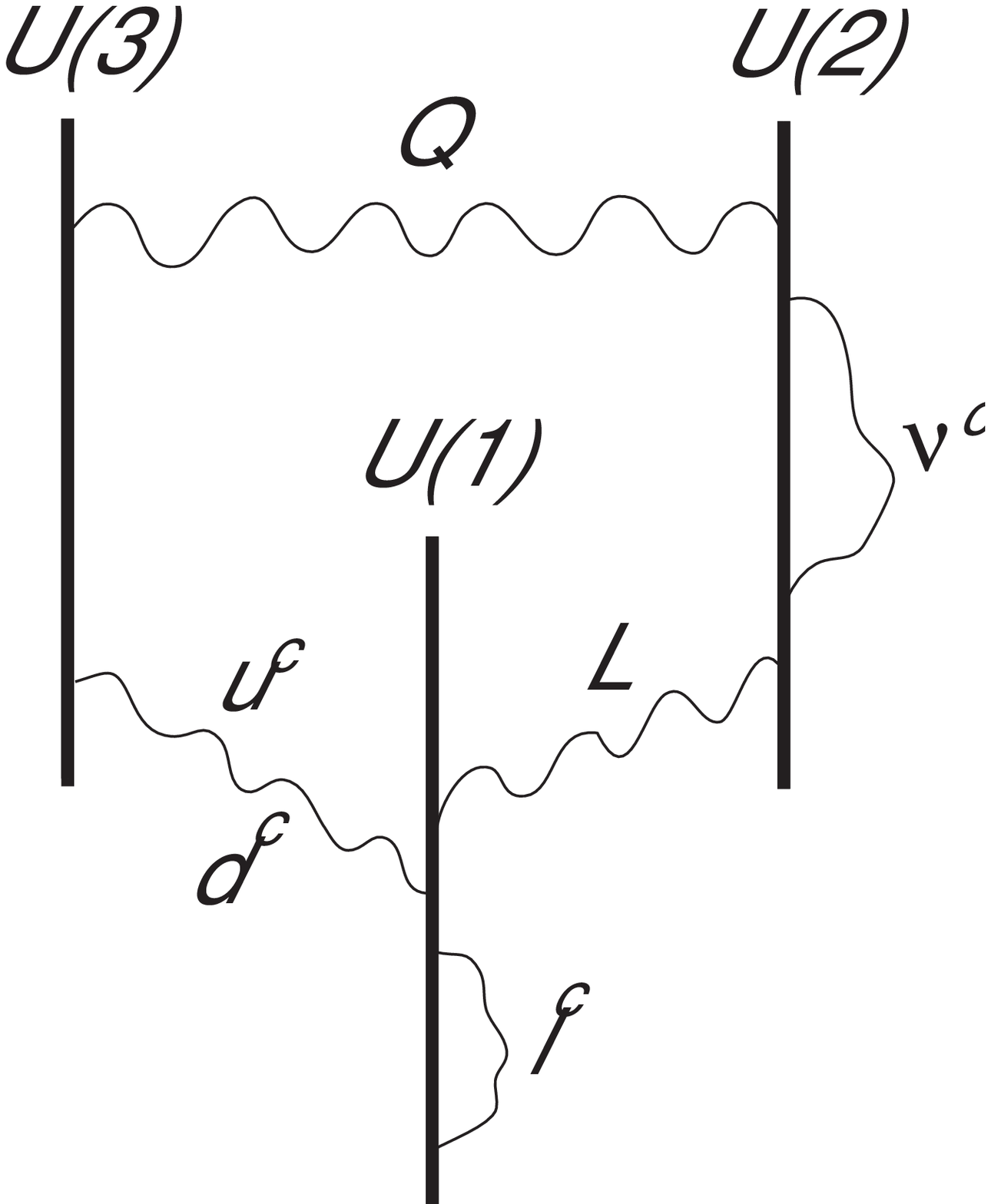}}
\caption{\label{fig_models}{\it Pictorial representation of models} $A$,  $B$ and $C$.}
\end{figure}

The model requires three stacks of branes giving rise to 
$U(3)\times U(2)\times U(1)$ gauge group. For completeness, 
below we will make a general study of SM embedding
in three brane stacks~\cite{ar}.
The quark and lepton doublets ($Q$ and $L$) correspond 
to open strings stretched between the weak and the color or $U(1)$ 
branes, respectively. On the other hand, the $u^c$ and $d^c$ antiquarks 
can come from strings that are
either stretched between the color and $U(1)$ branes, or that have 
both ends on the color branes and transform in the antisymmetric 
representation of $U(3)$ (which is an anti-triplet). There are 
therefore three possible models, depending on whether it is the $u^c$ 
(model A), or the $d^c$ (model B), or none of them (model C), the 
state coming from the antisymmetric representation of color branes. 
It follows that the antilepton $l^c$ comes in a similar way  from 
open strings with both ends either on the weak brane stack and 
transforming in the antisymmetric representation of $U(2)$ which is 
an $SU(2)$ singlet (in model A), or on the abelian brane 
and transforming in the ``symmetric" representation of $U(1)$
(in models B and C). The three
models are presented pictorially in Figure~\ref{fig_models}.

Thus, the members of a family of quarks and leptons have the 
following quantum numbers:
\ba
&&{\rm Model\ A}\qquad\qquad\quad\quad 
{\rm Model\ B}
\qquad\qquad\quad\ {\rm Model\ 
C}\nonumber\\
\!\!\!\!\!\!\!\! Q && ({\bf 3},{\bf 
2};1,1,0)_{1/6}\quad\quad\quad\ ({\bf 3},{\bf 2};1,\varepsilon_Q 
,0)_{1/6}
\qquad\, ({\bf 3},{\bf 2};1,\varepsilon_Q 
,0)_{1/6}\nonumber\\
\!\!\!\!\!\!\!\! u^c && (\bar{\bf 3},{\bf 
1};2,0,0)_{-2/3}\quad\quad\ \ \, (\bar{\bf 3},{\bf 
1};-1,0,1)_{-2/3}
\quad\ (\bar{\bf 3},{\bf 
1};-1,0,1)_{-2/3}\nonumber\\
\!\!\!\!\!\!\!\! d^c && (\bar{\bf 
3},{\bf 1};-1,0,\varepsilon_d )_{1/3}\qquad\, (\bar{\bf 3},{\bf 
1};2,0,0)_{1/3}
\qquad\ \ \, (\bar{\bf 3},{\bf 
1};-1,0,-1)_{1/3}
\label{model}\\
\!\!\!\!\!\!\!\! L && ({\bf 1},{\bf 
2};0,-1,\varepsilon_L)_{-1/2}\quad\ \, ({\bf 1},{\bf 
2};0,\varepsilon_L,1)_{-1/2}
\quad\ \ \, ({\bf 1},{\bf 
2};0,\varepsilon_L,1)_{-1/2}\nonumber\\
\!\!\!\!\!\!\!\! l^c && ({\bf 
1},{\bf 1};0,2,0)_1\qquad\quad\quad\, ({\bf 1},{\bf 
1};0,0,-2)_1
\qquad\ \ ({\bf 1},{\bf 
1};0,0,-2)_1\nonumber\\
\!\!\!\!\!\!\!\! \nu^c && ({\bf 1},{\bf 
1};0,0,2\varepsilon_\nu)_0\qquad\quad\ ({\bf 1},{\bf 
1};0,2\varepsilon_\nu,0)_0\qquad\ \, ({\bf 1},{\bf 
1};0,2\varepsilon_\nu,0)_0\nonumber
\ea
where the last three digits after the semi-column 
in the brackets are the charges under the three 
abelian factors $U(1)_3\times U(1)_2\times U(1)$, that we will call 
$Q_3$, $Q_2$ and $Q_1$ in the following, while the subscripts denote 
the corresponding hypercharges. The various sign ambiguities 
$\varepsilon_i=\pm 1$ are due to the fact that the corresponding 
abelian factor does not participate in the hypercharge combination 
(see below). 
In the last line, we also give the quantum numbers of a possible 
right-handed neutrino in each of the three models. These are in fact 
all possible ways of embedding the SM spectrum in three sets of 
branes.

The value of the weak angle can be easily computed from the 
hypercharge combination: 
\be
Y=\sum_ic_iQ_i\quad\Rightarrow\quad
\sin^2\theta_W={1\over 
1+4c_2^2+2c_1^2\alpha_2/\alpha_1+6c_3^2\alpha_2/\alpha_3}\, ,
\label{Ysin}
\ee
where $\alpha_i$ are the non-abelian couplings and the 
numerical coefficients are due to our normalization of $U(1)$ 
charges according to eq.~(\ref{gua}). In our models, the hypercharge 
combination is:
\ba
\label{hyper}
{\rm Model\ A}\quad\ &:&\quad Y=-{1\over 3}Q_3+{1\over 2}Q_2\\
{\rm Model\ B, C}&:&\quad 
Y=\ \ \, {1\over 6}Q_3-{1\over 2}Q_1\nonumber
\ea
leading to 
the following expressions for the weak angle:
\ba
{\rm Model\ A}\quad\ 
&:&\quad \sin^2\theta_W={1\over 2+2\alpha_2/3\alpha_3}
={3\over 8}\, {\bigg|}_{\alpha_{_2}=\alpha_{_3}}\\
{\rm Model\ B, C}&:&\quad 
\sin^2\theta_W={1\over 1+\alpha_2/2\alpha_1+\alpha_2/6\alpha_3}=
{6\over 
7+3\alpha_2/\alpha_1}\, 
{\bigg|}_{\alpha_{_2}=\alpha_{_3}}\nonumber
\label{sintheta}
\ea

In the second part of the above equalities, we used the unification relation 
$\alpha_2=\alpha_3$, that can be naturally imposed as described in 
section~\ref{unification}. Indeed, it follows by requiring the absence of 
chiral fermions that transform in the symmetric representation of $SU(3)$ 
and $SU(2)$ and the magnetic fields to be roughly an order of magnitude 
smaller than the string scale. The last condition can be partly relaxed 
in model A from the requirement of absence of chiral quark 
``anti"-doublets in the spectrum. Notice that such states have wrong 
hypercharge, since their $Q_2$ charge is opposite from quark
doublets in eq.~(\ref{hyper}).
This cannot be used in models B 
and C because $Q_2$ does not participate in the hypercharge 
combination, and thus, doublets and anti-doublets are 
indistinguishable. In any case, in these models the unification of 
the two non-abelian couplings is not sufficient to predict the weak 
angle and further conditions are needed for the $U(1)$ coupling 
$\alpha_1$. Such an analysis goes beyond the scope of this paper, 
which is to describe the general framework and present 
a simple example. Indeed, model A admits natural gauge 
coupling unification of strong and weak interactions, realizing the 
conditions we described in section~\ref{unification}, 
and predicts the correct value for 
$\sin^2\theta_W=3/8$ at the unification scale $M_{\rm GUT}$.

The spectrum (\ref{model}) can be easily implemented 
with a Higgs sector, since the Higgs field $H$ has the same 
quantum numbers as the lepton doublet or its complex conjugate:
\ba
&& {\rm Model\ 
A}\qquad\qquad\quad\quad {\rm Model\ B, C}
\nonumber\\
\!\!\!\!\!\!\!\! H\ && ({\bf 1},{\bf 
2};0,-1,\varepsilon_H)_{-1/2}\quad\ ({\bf 1},{\bf 2};0,\varepsilon_H,1)_{-1/2}
\\
\!\!\!\!\!\!\!\! H' && ({\bf 
1},{\bf 2};0,1,\varepsilon_{H'})_{1/2}\qquad\ \, 
({\bf 1},{\bf 
2};0,\varepsilon_{H'},-1)_{1/2}
\nonumber
\label{higgs}
\ea
Actually, as explained in the general framework of 
section~\ref{framework}, the Higgs sector should be locally 
supersymmetric, so that the Higgs scalars are massless
at the tree level,
and thus $H$ and $H'$ correspond to two Higgs chiral supermultiplets.

Besides the hypercharge combination, there are two additional 
$U(1)$s. It is easy to check that one of the two can be identified 
with $B-L$.  For instance, in model A choosing the signs 
$\varepsilon_d=\varepsilon_L=-\varepsilon_\nu=
-\varepsilon_H=\varepsilon_{H'}$, it is given by:
\be
B-L=-{1\over 6}Q_3+{1\over 2}Q_2-{\varepsilon_d\over 2}Q_1\, .
\label{BL}
\ee
The other $U(1)$ corresponds to a Peccei-Quinn (PQ) type symmetry. 
$B-L$ can be broken by a vacuum expectation value (VEV) of 
a SM singlet field of the type of $\nu^c$, at a high scale. 
In any case, this model has no baryon number conservation 
and thus proton is unstable by dimension six 
effective operators suppressed by the string scale.

The second $U(1)$ combination of PQ type is anomalous. 
The corresponding gauge field should become 
massive via the Green-Schwarz mechanism, by absorbing an axion from 
the Ramond-Ramond (RR) closed string sector~\cite{Sagnotti:1992qw}. 
Usually, its global counterpart survives and remains unbroken 
in perturbation theory at the orbifold point~\cite{Poppitz:1998dj}. 
To avoid the presence of an electroweak axion, one 
should either move away from this point, or 
find some appropriate extension of the model which allows to break
PQ by a scalar VEV at a high scale.
On the other hand, in the presence of magnetic fields, it was noticed
that the RR axions involved in the anomaly cancellation
come from the untwisted orbifold sector~\cite{Angelantonj:2000hi}.
In this case, the global symmetry will be in general broken at the scale
of the anomalous $U(1)$ mass, as in heterotic string models. 
As a result, the axion becomes invisible 
and no PQ symmetry should survive at low energies.

\section{Constraints from Gluino Cosmology} \label{pheno}

The most distinctive signature of split SUSY, decisively
differentiating it from the usual supersymmetric SM, 
is the long-lived gluino, which is the smoking gun of this framework.
 Because the scale of supersymmetry breaking is high,
the squarks are heavy and the lifetime for the gluino to decay
into a quark, antiquark and LSP -- which is mediated by virtual
squark exchange -- is:
\begin{equation}
\tau = 3 \times 10^{-2} {\rm sec} \Big(\frac{m_S}{10^9\,
{\rm GeV}}\Big)^4 \Big( \frac{1\, {\rm TeV}}{m_{\tilde{g}}}\Big)^5,
\end{equation}
where $m_S$ is the squark mass and $m_{\tilde{g}}$ the gluino mass.
We have included a QCD enhancement factor of $\sim 10$ in the
rate, and another factor $\sim 10$ for the number of decay
channels. The longevity of the gluino can lead to a host of interesting 
signatures at the LHC such as displaced vertices, intermittent tracks, 
late decaying gluinos captured near the detector etc., 
which have been discussed in refs.~\cite{savnim,noi,tutti,Arkani-Hamed:2004yi}. 
These signatures depend on the lifetime, which in turn depends 
sensitively, through the above equation, on the gluino mass and 
the squark mass $m_S \sim \sqrt{\delta H}$. These quantities are 
constrained by cosmological considerations, to which we turn next.

The most natural value for the squark mass, one that does not require 
tuning the ratio $\delta H/H$ to be small, is 
$m_S \sim \sqrt{\delta H} \sim M_{GUT} \sim 10^{16}$ GeV. 
For this value of $m_S$, and for  $m_{\tilde{g}} \sim 1$ TeV,
the gluino lifetime is of order $3 \times 10^{26}$ sec, much longer 
than the age of the universe. Such cosmologically stable gluinos are 
expected to assemble into color singlet ``R-hadrons" by combining 
with gluons, quarks and antiquarks during the QCD phase transition. 
Subsequently, during the primordial big bang nucleosynthesis, the 
R-hadrons are expected to assemble, often together with ordinary 
nucleons, into nuclei, which will eventually form atoms. 
These atoms will be chemically similar to a familiar atom, 
but will instead have a heavy $\sim$ TeV mass nucleus. 
Searches for such anomalous heavy isotopes are very restrictive. 
The limits on heavy hydrogen isotopes in the mass range up to 
$\sim$ 1 TeV is one such atom in $10^{27}$ nucleons, and for 
isotopes in the mass range from 1 TeV to 10 TeV is $10^{22}$ 
per nucleon~\cite{Smith:1982qu}. The upper limits for heavy (up to 10 TeV) 
isotopes of Helium, Carbon and Oxygen nuclei are 
as small as one atom in $\sim 10^{17}$ nucleons. 

These suggest similar limits to the abundances of gluinos relative to 
ordinary matter, since most gluinos and R-hadrons are expected to 
end-up in nuclei. Although one cannot prove that this is inescapable, 
it is hard to imagine that none of the many possible ways in which 
R-hadrons and ordinary hadrons  can combine with nucleons into 
some low-Z nuclei is realized. More precisely, there is a multitude 
of ways in multi-quark states can combine with a gluino into a 
color-singlet state of charge 0,1, 2, 6 or 8, and it is unlikely that none 
of these bound states form. So the abundance of gluinos relative to 
ordinary matter should probably be as small as $10^{-27}$ per nucleon, 
or $10^{-37}$ per photon, to account for the absence of heavy 
hydrogen isotopes of mass up to 1 TeV, or less than $10^{-32}$ per 
photon, if the gluino weighs up to 10 TeV. The absence of heavy, 
up to 10 TeV, isotopes of Helium, Carbon and Oxygen gives an 
upper limit of about  not bigger than  $10^{-17}$ per nucleon, 
or about $10^{-27}$ per photon. We now estimate the 
cosmological abundance of gluinos relative to photons 
before they have a chance to decay.

Before decaying, the gluinos can only reduce their number 
density by annihilating with each other. They can do so as either 
bare gluinos, before the QCD phase transition, or as gluinos 
clothed into R-hadrons, after the QCD phase transition. 
The cross section for bare gluinos is perturbative and 
scales as  $\sim m_{\tilde{g}}^{-2}$. The cross section for 
two R-hadrons to annihilate in the early universe is more 
subtle, and is still an open question; in part because even if the 
two R-hadrons combine into an ``R-molecule" bound state,
this can be dissociated by collisions with the medium before 
the gluinos in the molecule have a chance to annihilate each other. 
Making, nevertheless, the plausible hypothesis that the cross 
section for two R-hadron annihilation scales as the square of the QCD size,
of order $\sigma \sim 30$ mb,  results in a
gluino abundance which we estimate by equating expansion and
reaction rates, 
$n\, \sigma\, v \sim T^2/M_{Pl}$
with $T \sim \Lambda_{QCD}$. This translates to
\begin{equation}
\frac {n_{\tilde{g}}}{n_{\gamma}} = 10^{-18} \left(\frac
{m_{\tilde{g}}}{ 1\, \mbox{TeV}}\right)^{1/2}\, .
\label{eq:abundance1}
\end{equation}
This is larger than the the maximal allowed abundance of $10^{-27}$ 
relative to photons from the absence of anomalous heavy isotopes, 
and much larger than the upper limit of $10^{-37}$ ($10^{-32}$) 
per photon from the absence of anomalous isotopes of heavy 
hydrogen of mass up to 1 TeV ($10$ TeV). Though the estimate 
of the R-hadron annihilation cross section is uncertain, 
the discrepancy is so large that we can exclude the possibility 
of stable gluinos with some confidence. 

Some remaining loopholes are: 
1) The reheat temperature of the universe after inflation is 
so low that no gluinos are made. 
2) Gluinos do not form any heavy nuclei, which, as mentioned before, 
we find implausible. 
3) The simplest loophole: the gluino is in fact not cosmologically stable, 
and lives much less than the age of the universe. This can be 
accomplished most simply by a combination of a heavier gluino 
and lighter $m_S$. For example, if $m_S \sim \sqrt {\delta H} \leq 10^{15}$ GeV 
and the gluino mass is more than about $10$ TeV, 
the lifetime drops to $\sim 10^{16}$ 
years or less, which is acceptable because most gluinos will have 
decayed by now and will not be around to form heavy isotopes. 
Note that this requires some fine-tuning to make the ratio 
$\delta H/H$ small. This tuning though is radiatively stable
as was already pointed out in section \ref{susybr},
since it is protected by supersymmetry, which is broken by 
$\delta H$ but not $H$. Note however that for gluino masses 
much heavier than $\sim 10$ TeV, the successful gauge 
coupling unification will be distorted. In addition, such 
heavy gluinos will not be accessible to the LHC. 
A phenomenologically more appealing case is that of a 
TeV-mass gluino and $m_S \sim \sqrt {\delta H} \leq 3 \times 10^{13}$ GeV. 
This requires more of a tuning, which as before is radiatively stable, 
and maintains both unification and accessibility of gluinos at the LHC.

Gluinos, such as those just discussed, can also be cosmologically 
dangerous if their lifetime is shorter than the age of the universe 
but longer than a second, and their abundance is not adequately small. 
This is because their decay products can
distort the photon background or destroy nuclei synthesized during
primordial nucleosynthesis, which began when the universe was one
second old. A gluino that decays in less than a second is
harmless, as its decay products thermalize
and the heat bath erases any trace of its existence.
Gluinos that live longer than a second can be safe, as long as
their abundance is small. This is easily satisfied as long as the 
R-hadrons annihilate with a QCD-size cross section of order of 30 mb. 
The relevant quantity then, more important that the plain abundance, is:
\begin{equation}
m_{\tilde{g}}\, \frac {n_{\tilde{g}}}{n_{\gamma}} = 10^{-15}
\left(\frac {m_{\tilde{g}}}{ 1\,\mbox{TeV}}\right)^{3/2} \,
\mbox{GeV}\, .
\label{eq:abundance2}
\end{equation}
It measures the destructive power of the decaying gluino gas, 
as it depends on both the mass and the concentration of gluinos.
The abundance of gluinos with lifetime up to $10^{13}$ sec must be
small to  avoid  spectral distortions of the CMBR~\cite{Hu:1993gc}.
This constraint is mild, and equation (\ref{eq:abundance2}) easily
satisfies it. The abundance of gluinos with lifetime in the range
from $10^{-1}$ sec to $10^{12}$ sec must also be small to avoid
the destruction of the light nuclei synthesized during the
BBN~\cite{Dimopoulos:1988ue,Kawasaki:1994af}. Although this
constraint is strong, especially for lifetimes between $10^{4}$
sec to $10^{7}$ sec, equation (\ref{eq:abundance2}) satisfies it.
Other constraints from possible distortions of the diffuse photon
background  are also easily satisfied.

In summary, as long as its lifetime is much shorter than the age of the universe,
and the R-hadrons annihilate with QCD-size cross sections $\sim 30$ mb, 
the gluino is cosmologically safe, and does not distort either photon 
backgrounds or nuclear abundances. For a TeV mass gluino this entails 
a $\leq 10^{-3}$ fine-tuning that makes $\delta H$ smaller than $H$, 
and is protected by supersymmetry. Heavier gluinos require less tuning, 
at the expense  of distorting the successful unification 
and losing the gluinos at the LHC.

What if the R-hadrons  do not annihilate with QCD-size cross sections, 
and the only mechanism for the disappearance of gluinos before they 
decay is standard perturbative annihilation? Then to avoid distorting the 
photon spectrum or the nuclear abundances via the gluino decay products, 
its lifetime must be less than a second, which implies a squark mass 
$m_S \sim \sqrt {\delta H} \simlt 3 \times 10^9$ GeV, for a gluino mass 
of a TeV. Again, such a small $\delta H$ will require a tuning which is 
stable and protected by supersymmetry.

\section*{Acknowledgements}

We would like to thank Fabio Zwirner for useful discussions on the 
Scherk-Schwarz effective field theory. We also acknowledge valuable discussions
with Nima Arkani-Hamed, Gian Giudice and Marc Tuckmantel.
This work was supported in 
part by the European Commission under the RTN contracts 
HPRN-CT-2000-00148 and MRTN-CT-2004-503369. 
SD is supported by the NSF grant 0244728.

\setcounter{equation}{0}

\end{document}